\documentclass{jfm}
\usepackage{color}
\usepackage{amstext}
\usepackage{amssymb}
\usepackage{graphicx}
\usepackage{amsmath}

\newcommand\const{\mathrm{const}}

\begin{document}

{\title[Homogenization Method,  uncovered on 2 pages ] {Homogenization Method, \\ uncovered on 2 pages}}

\author[V. Vladimirov ]
{V.\ns A.\ns V\ls l\ls a\ls d\ls i\ls m\ls i\ls r\ls o\ls v}

\affiliation{DOMAS, Sultan Qaboos University, Oman and DAMTP, University of Cambridge, UK}

\date{December 5th 2014}

\setcounter{page}{1}\maketitle \thispagestyle{empty}

\begin{abstract}
This note gives a brief and `crash' introduction to the method of Homogenization with the use of wave equation and  diffusion equation with periodic in space coefficients as instructive examples.
We expose the method with the use of an approach, which appears in Vibrodynamics. The note can be interesting for people who want to use  the Homogenization method immediately.

\end{abstract}

\section{Introduction \label{sect01}}

Homogenization represents a mathematical method actively used in high impact interdisciplinary research.
There are hundreds of related publications, even the list of recently published monographs contains about twenty titles, including \cite{Hom1, Hom2, Hom3, Hom4, Hom5, Hom6, Hom7,Hom8}. This literature shows that the using of Homogenization theory often requires rather advanced mathematical techniques, such as functional analysis, etc.
The aim of this note is to give a simplest possible `crash' introduction to the method of Homogenization by showing how it works for the description of waves in an elastic medium with periodically changing properties.
The exposition follows to systematic and justifiable procedure that was introduced in Vibrodynamics, see \cite{Vladimirov0,Yudovich, Vladimirov}.
The results include the deriving of equation for an averaged part of solution and an explicit formula for its oscillating part.
As a result, the equation for the averaged solution appears as the wave equation with an effective coefficient of elasticity, which is a well-known and classical result.
Transparency, generality, and simplicity are the main benefits of our approach.

\section{Formulation of problem and the two-scale setting}

We consider one-dimensional elastic medium with the density and elasticity properties periodically changing in the $x$-direction.
The wave equation for a longitudinal displacement $ {a} = {a} (x,t)$ is written in the dimensional form as
\begin{equation}\label{wave-eqn}
\boxed{\rho {a} _{tt}=(E {a} _x)_x;\quad \rho(x)=\rho(x+2\pi l), \quad E(x)=E(x+2\pi l)}
\end{equation}
where $t$ - time, subscripts stand for partial derivatives, $E>0$ -- Yong's modulus, $\rho>0$ -- mass per unit length, $2\pi l=\const >0$ -- period in $x$.
For functions $\rho$ and $E$  we take
\begin{equation}\label{coeff}
 \rho=\rho_0+\widetilde{\rho}>0,\quad E=E_0+\widetilde{E}>0
\end{equation}
where $\rho_0$, $E_0$ are constants, and $\widetilde{\rho}$, $\widetilde{E}$ are $2\pi l$-periodic in $x$ functions with zero average.
For example, $ \rho=\rho_0+\rho_1\cos( x/l),\quad E=E_0+E_1\cos( x/l)$
%
with constants  $\rho_1, E_1$.
We introduce the
dimensionless variables and parameters (marked with `dags') as follows:
\begin{equation}\nonumber\label{dimensionless}
   x=l x^\dag,  \quad t=T t^{\dag},\quad E =E_0 E^\dag,\quad \rho=\rho_0 \rho^\dag;\quad T \equiv\lambda\sqrt{\rho_0/ E_0}
\end{equation}
The dimensionless form of \eqref{wave-eqn} is (the `dags' are dropped):
\begin{eqnarray}\label{wave-eqn-dless}
&&\boxed{(E {a} _x)_x=  \varepsilon^2\rho {a} _{tt};
\quad\varepsilon\equiv l/\lambda,\quad \rho(x)=\rho(x+2\pi), \quad E(x)=E(x+2\pi)}
\end{eqnarray}
We accept that $\varepsilon$ is a small parameter $\varepsilon\equiv{l}/{\lambda} \ll 1$,
which means that we consider the large-scale solutions (of characteristic scale $\lambda\gg l$).

To solve (\ref{wave-eqn-dless}) we introduce \emph{two mutually dependent} spatial variables $\xi$ and $r$
\begin{equation}\nonumber\label{2scales}
 \boxed{ \xi=x,\quad r=\varepsilon x}
\end{equation}
and the form of the unknown function which depends on both variables
\begin{equation}\nonumber\label{2scales-1}
   {a} (x,t)={b}(\xi,r,t)
\end{equation}
With the use of the chain rule $\partial/\partial x=\partial/\partial \xi+\varepsilon \partial/\partial r$
%
the equation \eqref{wave-eqn-dless} takes the form
\begin{equation}\label{eqn-1}
(E{b}_\xi)_\xi+\varepsilon[(E{b}_\xi)_r+(E{b}_r)_\xi]+\varepsilon^2[(E{b}_r)_r-\rho{b}_{tt}]=0
\end{equation}
The key `multi-scale' or `two-scale' assumption is:
\begin{equation}\nonumber\label{key}
\boxed{\emph{We consider $\xi$ and $r$ as two mutually independent variables}}
\end{equation}

Hence we are looking for solutions ${b}(\xi,r,t)$ of the PDE \eqref{eqn-1} with three independent variables $t$, $\xi$ and $r$.
We express the solutions to  \eqref{eqn-1} as regular series
\begin{equation}\label{series}
{b}={b}_0+\varepsilon{b}_1+\varepsilon^2{b}_2+\dots,\quad {b}_i={b}_i(\xi,r,t)=O(1),\ i=1,2,3, \dots
\end{equation}
For  performing of asymptotic procedure we assume that all dimensionless function $b(\xi,r,t)$, $b_i(\xi,r,t)$ (say, denoted as $f$) has the following properties:

\noindent
$\bullet$  $f\sim {O}(1)$ and  all its $\xi$-, $r$-, and $t$-derivatives are also ${O}(1)$;

\noindent
$\bullet$ $f$  is $2\pi$-periodic in $\xi$, \emph{i.e.}\ $f(\xi, s, \tau)=f(\xi+2\pi,r,t)$;

\noindent
$\bullet$  $f$ has an average given by
\begin{equation}\label{average}
\langle {f}\,\rangle \equiv \frac{1}{2\pi}\int_{\xi_0}^{\xi_0+2\pi}
f(\xi, r, t)\, d \xi\equiv \overline{f}(r,t) \qquad \forall\ \xi_0=\const;
\end{equation}

\noindent
 $\bullet$  $f$ can be split into the average and purely oscillating parts
 $f(\xi, r, t)=\overline{f}(r, t)+\widetilde{f}(\xi, r, t),$
 where  \emph{tilde-functions} (or  purely oscillating functions) are such that $\langle \widetilde f\, \rangle =0$ and the \emph{bar-functions} (or averaged functions) $\langle f\rangle\equiv\overline{f}$ are $\xi$-independent.

 \section{Asymptotic Procedure}
 Let us describe the procedure of solving \eqref{eqn-1}, \eqref{series} for the first three successive approximations:

\textbf{\emph{The equation of order $\varepsilon^0$ is}}
\begin{eqnarray}
 && (E\widetilde{{b}}_{0\xi})_\xi=0\quad \text{or}\quad E\widetilde{{b}}_{0\xi}=\overline{F}_1(r,t) \label{zero}
 \end{eqnarray}
 where $\overline{F}_1(r,t)$ is an arbitrary function, which appears after integration over $\xi$. It is convenient to rewrite \eqref{zero} as
 \begin{eqnarray}\nonumber
 &&  \widetilde{{b}}_{0\xi}=M\overline{F}_1(r,t);\quad \text{where}\quad M\equiv 1/E, \quad E>0 \label{zero-a}
 \end{eqnarray}
Its bar-part is $\overline{M}\,\overline{F}_1=0$,
which immediately yields $\overline{F}_1=0$. Then  \eqref{zero} gives
\begin{eqnarray}
&&  \boxed{\widetilde{{b}}_{0\xi}=0\quad\text{or}\quad b_0=\overline{b}_0(r,t)}\label{zero-c}
\end{eqnarray}
with unknown function $\overline{b}_0(r,t)$. Surprisingly, the solution in  main approximation is not oscillating.

\textbf{\emph{The equation of order $\varepsilon^1$ is}}
\begin{eqnarray}\nonumber
&&(E\widetilde{{b}}_{1\xi})_\xi+(E{b}_{0\xi})_r+(E{b}_{0r})_\xi=0\label{one}
\end{eqnarray}
The use of \eqref{zero-c} simplifies it to the form $$(E\widetilde{{b}}_{1\xi}+E{\overline{b}}_{0r})_\xi=0$$
which after integration gives
\begin{eqnarray}
&&\widetilde{{b}}_{1\xi}+{\overline{b}}_{0r}=M\overline{F}_2(r,t)\label{one-aa}
\end{eqnarray}
with an arbitrary function $\overline{F}_2(r,t)$. Its bar-part yields $\overline{F}_2(r,t)={\overline{b}}_{0r}/\overline{M}$,
then the tilde-part of \eqref{one-aa} leads to
\begin{eqnarray}
&&\boxed{\widetilde{{b}}_{1\xi}={\overline{b}}_{0r}\widetilde{M}/\overline{M}}\label{one-aaaa}
\end{eqnarray}
This formula is the only result obtainable from the equation of the first approximation.

\textbf{\emph{The equation of order $\varepsilon^2$ is}}
\begin{eqnarray}\nonumber
&&(E\widetilde{{b}}_{2\xi})_\xi+(E{b}_{1r})_\xi+(E\widetilde{{b}}_{1\xi})_r+(E{b}_{0r})_r-\rho {b}_{0tt}=0\label{second}
\end{eqnarray}
its bar-part is
\begin{eqnarray}
&&\langle \widetilde{E}\widetilde{{b}}_{1\xi}\rangle_r+(\overline{E}\,\overline{{b}}_{0r})_r-\overline{\rho} \overline{{b}}_{0tt}=0\quad \nonumber
\end{eqnarray}
which jointly with \eqref{one-aaaa} leads to:
\begin{eqnarray}\label{second-a}
\overline{\rho}\, \overline{{b}}_{0tt}=[(\overline{E}+\langle\widetilde{E}\,\widetilde{M}\rangle/\overline{M})\,\overline{{b}}_{0r}]_r
\end{eqnarray}
The substitution $\widetilde{E}=E-\overline{E}$, $\widetilde{M}=M-\overline{M}$ yields
$
\overline{E}+\langle\widetilde{E}\,\widetilde{M}\rangle/\overline{M}=1/\overline{M}
$
and the final form of \eqref{second-a} is
\begin{eqnarray}\label{main-eq}
\boxed{\overline{\rho}\, \overline{{b}}_{0tt}=(E_\text{eff}\,\overline{{b}}_{0r})_r,\quad\text{where}\quad E_\text{eff}=\langle E^{-1}\rangle^{-1}}
\end{eqnarray}
where the calculation of average $\langle\cdot\rangle$ represents the integration over $\xi$ \eqref{average}. 

Hence, we have derived the wave equation for the solution in the main approximation,  with an effective Young's module $E_\text{eff}=\const$. One can notice, that we haven't used any assumption for the smallness of oscillation parts in the given functions $E$ and $\rho$ \eqref{coeff}.
Similar results represent the main and typical outcomes of all Homogenization approaches. One can also notice, that in addition to the equation \eqref{main-eq} for the averaged part $\overline{{b}}_{0}$ of the solution we have also obtained the explicit formula for its oscillating part $\widetilde{b}_1$ \eqref{one-aaaa}.

\section{Discussion}

In all the above consideration we have exploited only the $x$-derivatives in the equation, and have left the time derivative in its original form.   Hence, it is possible to consider any other equation (PDE) containing the same $x$-derivatives (and different time-derivatives) applying the same two-scale procedure as above, to have the same outcome for the `spatial' part of the equation. The best example to demonstrate this statement is to consider the diffusion equation
\begin{eqnarray}
a_t=(K a_{x})_x, \qquad K(x)=K_0+\widetilde{K}
 \end{eqnarray}
where $a=a(x,t)$ -- is a passive admixture or temperature, $K_0$ is constant, $K=K(x)>0$ -- the diffusion coefficient  which is assumed to be a periodic function $K(x)=K(x+2\pi)$.
The procedure is identical to the given above and the expression for the effective diffusion coefficient is similar to $E_\text{eff}$ in \eqref{main-eq}. Hence, the averaged diffusion equation is:
\begin{eqnarray}
\overline{a}_{0t}=(K_\text{eff} \overline{a}_{0r})_r, \quad K_\text{eff}=\langle K^{-1}\rangle^{-1}
\end{eqnarray}
Interesting examples showing that the exposed method is highly efficient and useful can be easily continued.


\begin{acknowledgments}
This introduction to the method of Homogenization represents a part of lectures delivered by the author in the Colorado State University in 2003 and 2004 and in the Tsinghua University and Peking University in 2007. The author is grateful to Prof. M.T. Montgomery and Prof. Din-Yu Hsieh for organizing these lectures. This particular text was prepared as a supporting material for the research seminars in Vibrodynamics in SQU, Oman in 2015.
\end{acknowledgments}

\end{document}